\documentclass[prb,twocolumn,a4paper,showpacs,showkeys,floatfix]{revtex4}
\usepackage{graphicx}	
\usepackage{amsmath}	
\usepackage{dcolumn}	
\newcommand{\me}{\text{e}}	

\begin{document}
\title{Intervalley splitting and intersubband transitions in n-type Si/SiGe quantum wells: pseudopotential vs. effective mass calculation}
\author{A. Valavanis}
\email{a.valavanis05@leeds.ac.uk}
\author{Z. Ikoni\'{c}}
\author{R. W. Kelsall}
\affiliation{Institute of Microwaves and Photonics, School of Electronic and Electrical Engineering, University of Leeds, Leeds LS2 9JT, United Kingdom}
\date{\today}

\begin{abstract}
Intervalley mixing between conduction band states in low-dimensional Si/SiGe heterostructures induces splitting between nominally degenerate energy levels.  The symmetric double-valley effective mass approximation (DVEMA) and the empirical pseudopotential method (EPM) are used to find the electronic states in different types of quantum wells.  A reasonably good agreement between the two methods is found, with the former being much faster computationally.  Aside from being an oscillatory function of well width, the splitting is found to be almost independent of in-plane wave vector, and an increasing function of the magnitude of interface gradient.  Whilst the model is defined for symmetric envelope potentials, it is shown to remain reasonably accurate for slightly asymmetric structures such as a double quantum well, making it acceptable for simulation of multilayer intersubband optical devices.  Intersubband optical transitions are investigated under both approximations and it is shown that in most cases valley splitting causes linewidth broadening, although under extreme conditions, transition line doublets may result.
\end{abstract}

\maketitle

\section{Theoretical considerations}
\subsection{Intervalley Mixing and the Effective Mass Approximation}
The conventional single-valley effective mass/envelope function approximation (EMA) is computationally efficient and remains remarkably accurate at the atomic scale.\cite{book:Harrison2005}  The behaviour of carriers in a 2-dimensional heterostructure is modeled by introducing an \emph{effective mass}, $m^*(z)$ and an \emph{envelope function} potential, $V(z)$ which depends on the material composition \cite{book:Davies98}.  The band non-parabolicity is optionally incorporated into the effective mass, via $m^*(z) = m_0\left[1+\alpha{}(E-V(z))\right]$. The Hamiltonian for such a system is:
\begin{equation}
H(z) = - \frac{\hbar^2}{2}\frac{\partial}{\partial{}z}\frac{1}{m^*(z)}\frac{\partial}{\partial{}z}+V(z)
\end{equation}

If there are multiple equivalent valleys in $k$-space, these must be considered independently within the ``pure'' EMA, with any intervalley mixing effects neglected.  The conduction band edge is located in the six $\Delta$-valleys for Si$_{1-x}$Ge$_x$ alloys with $x<85\%$ \cite{article:SSTPaul04}.  Strain effects split the degenerate valleys into four $\Delta_{\parallel}$-valleys and two $\Delta_{\perp}$-valleys. As the $\Delta$-valleys are ellipsoidal, a different effective mass is used for each degenerate set.

With two equivalent $\Delta_{\perp}$-valleys, their quantum confinement subbands are predicted by the envelope function/EMA to be degenerate.  In reality, there is a coupling between the two sets of states, which results
in a splitting, i.e. lifting of the degeneracy.  This effect has been observed experimentally in Shubnikov-de Haas oscillation measurements in high magnetic fields \cite{article:SurfSciWeitz1996, article:SSTKoester1997, article:SSCKoehler1978, article:SSCNicholas1980, article:PRFang1968, article:PRBStoger1994, article:PRBKhrapai2003, article:PRLLai2004} with energy splitting up to a few meV.  Boykin \emph{et al} presented a tight-binding model of the ground state splitting in a biased square quantum well with both hard-wall and cyclic boundary conditions \cite{article:APLBoykin2004, article:EJPBoykin2005, article:PRBBoykin2004}.  The ground state splitting in an unbiased square well was shown to be approximated by:
\begin{equation}
\label{eqn:BoykinSplitting}
\Delta{}E_1\approx\frac{16\pi^2u}{(S+2)^3}\sin\left(
\frac{\phi_{\text{min}}}{2}\right)\left|\sin\left[(S+2)
\frac{\phi_{\text{min}}}{2}\right]\right|
\end{equation}
where $\phi_{\text{min}}=k_0a$, and $k_0$ denotes the position of the valley minimum in the Brillouin zone, $a$ is the lattice constant, $S$ is the number of crystalline monolayers in the quantum well and $u$ is a fitting constant.  From this model, it is clear that the ground state splitting oscillates with well width; the frequency being dependent on the location of the valley minima.  Similar results have been obtained for the two lowest subbands in an unbiased well by Chiang \cite{article:JJAPChiang1994} in an anti-bonding orbital model, and recently by Nestoklon \cite{article:PRBNestoklon2006} in a slightly different tight-binding model.

Splitting due to an electric field has been considered by modelling a triangular quantum well. The composition profile of the structure in this case is selected to provide a potential gradient \emph{i.e.} an internal electric field.  Although an effective mass model by Sham \cite{article:PRBSham1979} proposed that the splitting is simply proportional to the applied field, Boykin \emph{et al} \cite{article:JAPBoykin2005} and Grosso \emph{et al} \cite{article:PRBGrosso1996} show that the splitting is a non-linear function of both the well width and the electric field.

Modifications to the EMA have been introduced in order to describe the intervalley mixing effects in an infinite square well \cite{article:JPSJOhkawa1977_1, article:JPSJOhkawa1977_2, article:SSCOhkawa1978} and a finite square well with impurity states \cite{article:PRBFang2005}.  Ting and Chang's Double Valley Effective Mass Approximation (DVEMA) \cite{article:PRBTing1988} provides an elegant self-contained EMA description of intervalley mixing effects.  This model allows a computationally fast simulation of long structures such as Quantum Cascade Lasers (QCLs)
\cite{article:RepProgPhysGmachl2001}, where the use of atomistic methods would be extremely cumbersome.  The model is applicable to any symmetric conduction band-edge envelope potential, $\hat{V}(z)$.  The kinetic energy operator in the Hamiltonian is unchanged, but a \emph{splitting potential} $\hat{U}(z)$ is included to account for intervalley mixing:
\begin{equation}
\hat{H}(z)=-\frac{\hbar^2}{2}\frac{\partial}{\partial{}z}\left(\frac{1}{m(z)}
\frac{\partial}{\partial{}z}\right)+\hat{V}(z)\pm{}\hat{U}(z).
\end{equation}
The derivation of Ting and Chang's DVEMA \cite{article:PRBTing1988} starts with the assertion that in $k$-space there are two equivalent $\Delta_{\perp}$-valleys at wave-vectors, $k\pm{}k_0$.  By symmetry, the wavefunction can be decomposed into either even or odd symmetric combinations of functions centred at the valley minima with equal and real weighting coefficients:
\begin{equation}
\label{eqn:basis}
\left|k_{1,2}\right>=\frac{1}{\sqrt{2}}
\left(\left|k+k_0\right>\pm\left|k-k_0\right>\right).
\end{equation}
In the most general case, the weighting factors of the two terms may become complex conjugates, with magnitude $\frac{1}{\sqrt{2}}$ \cite{unpublished:Friesen2006}.  The complete wavefunction is now defined as:
\begin{equation}
\left|\psi_{1,2}\right>=\sum_k{}\phi_{1,2}(k)\left|k_{1,2}\right>.
\end{equation}
The matrix elements of the envelope potential operator in the Hamiltonian are written in the basis defined by Eq.~(\ref{eqn:basis}):
\begin{equation}
\begin{split}
V_{nm}=\left<k_n\right|\hat{V}\left|k_m\right>&=\frac{1}{2}\left(\left<k_n+k_0\right|\pm\left<k_n-k_0\right|\right)\\
&\times\hat{V}\left(\left|k_m+k_0\right>\pm\left|k_m-k_0\right>\right)
\end{split}
\end{equation}
Rearranging this expression yields:
\begin{equation}
\begin{split}
V_{nm}&=\frac{1}{2}\left(\left<k_n+k_0\right|V\left|k_m+k_0\right>+
\left<k_n-k_0\right|V\left|k_m-k_0\right>\right)\\
&\pm\frac{1}{2}\left(\left<k_n+k_0\right|V\left|k_m-k_0\right>+
\left<k_n-k_0\right|V\left|k_m+k_0\right>\right)
\end{split}
\end{equation}
Using the discrete to continuous approximation, $V_{nm}=\tilde{V}(k_m-k_n)$, the intervalley envelope term can be written as:
\begin{equation}
\tilde{V}_{1,2}(k)=\tilde{V}\left(k\right)\pm\frac{1}{2}\left[
\tilde{V}\left(k-2k_0\right)+\tilde{V}\left(k+2k_0\right)\right]
\end{equation}
where $\tilde{V}(k)$ is the Fourier transform of the conduction band-edge envelope potential.  The configuration-space form of the intervalley envelope function is found by taking the inverse Fourier transform of this result and the splitting potential is therefore extracted as:
\begin{equation}
\label{eqn:SplittingPotential}
\begin{split}
\hat{U}(z)&=\frac{1}{2}\mathcal{F}^{-1}\left\{\tilde{V}\left(k-2k_0\right)+
\tilde{V}\left(k+2k_0\right)\right\}\\
&=V(z)\cos(2k_0z)
\end{split}
\end{equation}
where $\mathcal{F}^{-1}$ denotes the inverse Fourier transform.  This form of real-space splitting potential is more general than the special case of a square well with abrupt interfaces \cite{article:PRBTing1988}.  It applies to any symmetric potential, but does not explicitly show that the splitting potential oscillates as a function of well width $W$ of the form $\sin(k_0W)/k_0$ \cite{article:PRBTing1988}.  In addition, the numerical solution is complicated by the presence of a continuous, rapidly varying splitting function, as opposed to a pair of delta-functions at the square well interfaces \cite{article:PRBTing1988}.  However, since the problem is one-dimensional, this is not a major concern.  For asymmetric envelope functions, the simple cosine form is no longer strictly valid as the Fourier coefficients in equation \ref{eqn:basis} become a complex conjugate pair.  However, it will be shown here that the symmetric approximation still gives excellent results for structures with a moderate degree of asymmetry.

\subsection{Empirical Pseudopotential Calculation}
\begin{table*}[tb]
\caption{\label{tbl:Pseudopotential}Pseudopotential parameters\cite{article:PRBFriedel1989}.}
\begin{ruledtabular}
\begin{tabular}{ldddddd}
Parameter&a_1&a_2&a_3&a_4&a_5&a_6\\
\hline 
Si&212.1372&2.2278&0.6060&-1.9720&5.0&0.3\\
Ge&108.9024&2.3592&0.7400&-0.3800&5.0&0.3
\end{tabular}
\end{ruledtabular}
\end{table*}

An empirical pseudopotential method (EPM) is used here to calculate electronic states in Si/SiGe based quantum well structures, and provide a comparison with the results of the DVEMA simulation.  Being a microscopic method (in common with tight-binding), it normally reveals intervalley interference induced splitting of size-quantised subbands, without any adjustable parameters introduced for this purpose.  The supercell implementation of the EPM was used, with a continuous atomic formfunction, $V(g)$.  The ``modified Falicov'' formfunction described by
Friedel \emph{et al} \cite{article:PRBFriedel1989} was selected:
\begin{equation}
V(g)=\frac{a_1\left(g^2-a_2\right)}{1+\me^{\left[a_3\left(g^2-a_4\right)
\right]}}\cdot\frac{1}{2}\left[\tanh\left(\frac{a_5-g^2}{a_6}\right)+1\right]
\end{equation}
This formfunction has been used by Fischetti and Laux \cite{article:JAPFischetti1996}, as well as in previous work by the authors \cite{article:PRBIkonic2001_2}.  It gives reasonable agreement with experimental data for both bulk Si and Ge band structure and for band discontinuities at the interface.  A cut-off energy of 4.5\,Ry was used, which gives an acceptable number of plane waves for accurate and rapid computation with all the structures considered.  The parameters for Si and Ge are given in Table \ref{tbl:Pseudopotential}, and the virtual crystal approximation was used for the alloy.

The EPM can be used for structures with either abrupt interfaces or graded compositions.  In the latter case, the interface grading is piecewise constant (\emph{i.e.} within the crystalline monolayer width).  Individual layers are given the required alloy compositions.  In contrast, the DVEMA uses a continuous potential profile, though features smaller than the width of a monolayer have little practical meaning.

It is important to note that effective mass based calculations (like the DVEMA) can never fully reproduce the results
of microscopic EPM modelling.  This is because the DVEMA only handles four bulk states explicitly, while the EPM implicitly includes many evanescent states, stemming from remote bulk bands.  Furthermore, the location of indirect valleys may vary between the bulk materials used in the well and barrier --- a situation which is difficult to handle with effective mass methods.  Given that remote bands are usually less important than the bands from which the quantised states are derived, one can expect \emph{reasonable} accuracy from the DVEMA.  This may be validated by comparison against the EPM calculation.

\section{Numerical Results and Discussion}
DVEMA and EPM calculations were performed for a range of Si/SiGe Quantum Wells (QW), using the material parameters as follows. The lattice constant $a(x)$ of relaxed Si$_x$Ge$_{1-x}$ alloy is found by interpolation of the elemental lattice constants, given in table \ref{tbl:MaterialConstants} according to $a(x)=a_\text{Si}(1-x)+a_\text{Ge}x-b_\text{bow}x(1-x)$ \cite{article:SSTPaul04}, with the bowing parameter
$b_\text{bow}=0.2733\,\text{pm}$ \cite{article:PhysStatSolBublik1974}.  In a structure coherently grown in the [001] direction, comprising layers of different compositions on a relaxed virtual substrate, the strain results in
the perpendicular lattice constant of a particular layer being equal to $a_{\perp}(x)=a(x)\left[1-\left(2C_{12}(x)/C_{11}(x)\right)\left((a_0(x) - a(x))/a(x)\right)\right]$, while the in-plane lattice constants equal that of the substrate.  The elastic constants for SiGe are found by linear
interpolation of the values given in Table \ref{tbl:MaterialConstants}, and $a_0(x)$ is the bulk lattice constant of the relaxed virtual substrate on which the structure is grown.

In this work, the Ge fraction in the virtual substrate is fixed at 20\%.  For the DVEMA calculation, the $\Delta$-valley longitudinal and transverse effective masses are $m_L=0.916$ and $m_T=0.19$ respectively, and do not depend on the alloy composition.  In the EPM calculations the total length of the structure (\emph{i.e.} the supercell period) which includes the well and barrier layers, was set to a fixed value of 35 monolayers (ML), where 1\,ML is half the lattice constant (\emph{i.e.} twice the crystalline monolayer width of eqn. \ref{eqn:BoykinSplitting}).  This maintained a constant number of plane-waves in the pseudopotential basis set, and avoided fluctuations in the results caused by variable size of basis.  This is important since intervalley splitting is relatively small on the energy scale covered by EPM \cite{article:PRHensel1965}.

\begin{table}[tb]
\caption{\label{tbl:MaterialConstants}Material parameters for silicon and germanium}
\begin{ruledtabular}
\begin{tabular}{lddl}
Parameter&\text{Si}&\text{Ge}&Unit\\
\hline
$a$&543.1\footnotemark[1]&563.3\footnotemark[1]&pm\\
$C_{11}$&167.5\footnotemark[2]&131.5\footnotemark[3]&GPa\\
$C_{12}$&65.0\footnotemark[2]&49.4\footnotemark[3]&GPa
\footnotetext[1]{Reference \onlinecite{article:PhysStatSolBublik1974}.}
\footnotetext[2]{Reference \onlinecite{article:JAPMcSkimin1953}.}
\footnotetext[3]{Reference \onlinecite{article:JAPMcSkimin1964}.}
\end{tabular}
\end{ruledtabular}
\end{table}
\subsection{\label{scn:FiniteSquareWell}Finite Square Well}
\begin{figure}[tb]
\includegraphics*[width=0.8\columnwidth]{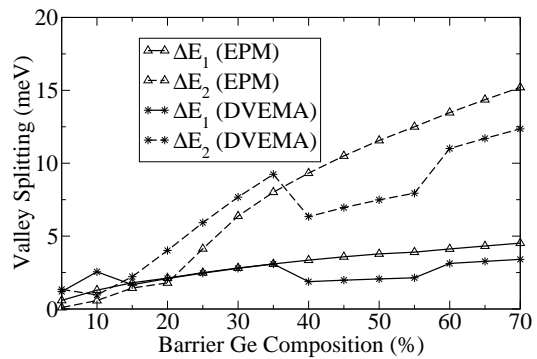}
\caption{\label{fig:splittingBarrierCompBoth}Intervalley splitting, $\Delta$E in the lowest two subbands of an 8\,ML quantum well as a function of barrier composition.}
\end{figure}

The first set of calculations was for a simple square QW with abrupt interfaces.  Fig.~\ref{fig:splittingBarrierCompBoth} shows the influence of barrier composition (potential height) on the splitting of the lowest two subbands, obtained by both the DVEMA and EPM calculations, for a fixed, 8\,ML wide Si quantum well.  The confining potential of the quantum well increases almost linearly with the Ge content in the barriers \cite{article:PRBRieger1993}.  The results show a clear increase in valley splitting with increasing confining potential, with the two sets of results being broadly in good agreement, and with the most obvious discrepancy being the discontinuities in the DVEMA plot.

\begin{figure}[tb]
\includegraphics*[width=0.8\columnwidth]{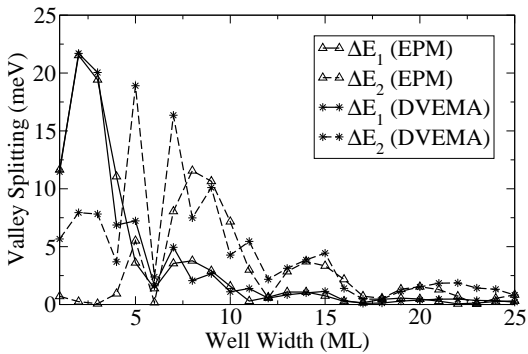}
\caption{\label{fig:splittingSquareBoth}Intervalley splitting in a Si/Si$_{0.5}$Ge$_{0.5}$ quantum well as a function of well width}
\end{figure}

The effect of well width upon state splitting was investigated next, for a structure with a fixed barrier composition of 50\% Ge.  The well width was varied between 1\,ML and 25\,ML in a supercell of total length 35\,ML.  This leaves a minimum 10\,ML barrier region, thus ensuring decoupling of neighbouring quantum wells under periodic boundary conditions.  It also represents the realistic range of well widths within intersubband optical devices.  Figure \ref{fig:splittingSquareBoth} shows the EPM and DVEMA results.  As predicted by equation \ref{eqn:BoykinSplitting}, the valley splitting is a decaying oscillatory function of well width, originating from interference of the wavefunction components reflecting at the quantum well interfaces.  Since the number of monolayers in the structure is restricted to the set of positive integers, the splitting function is undersampled to show the precise value of the period of oscillations.  DVEMA shows good agreement with the EPM results for the envelope of the splitting, whilst the oscillatory component is approximately correct.  For very small well widths however, the results for higher subbands deviate from the theory, as the states are no longer bound in the well.

The tight-binding results presented by Boykin \cite{article:APLBoykin2004} consider wide quantum wells, with low-energy ground state and effectively infinite barriers.  The decaying oscillatory form is however, the same as that obtained from DVEMA calculations.  Equation \ref{eqn:BoykinSplitting}, extracted from the tight-binding model, predicts a period of around 6\,ML as observed in fig.~\ref{fig:splittingSquareBoth}.  Setting $u\approx3$ yields a match in the amplitude between the three models.  This figure is somewhat higher than the value given in the reference above, although  equation \ref{eqn:BoykinSplitting} has been fitted to the DVEMA results for loosely bound states in a finite square well as opposed to being derived from bulk dispersion characteristics.

\subsection{The influence of in-plane Wave vector}
For nonzero in-plane wave vector $k_{\parallel}$, the size-quantised subbands in QWs generally acquire their $k_{\parallel}$-dependent contributions from remote bulk bands.  For bulk materials this is manifested as non-parabolicity of the in-plane dispersion, and in the problem under consideration, $k_{\parallel}$-dependent splitting would result.  The EPM calculation automatically accounts for these effects, while the DVEMA would require $k_{\parallel}$-dependent ``correction'' terms as remote bands are not explicitly included in the model.  The in-plane non-parabolicity of the $\Delta_{\perp}$-valley is relatively low, and indeed a rather weak dependence
of splitting on $k_{\parallel}$ is found.  In the EPM calculation for a 10\,ML quantum well, the splitting in the first and second subbands increased approximately linearly by 17\% and 8\% respectively, when $k_{\parallel}$ changed from zero to 10\% of the Brillouin zone edge (i.e. in the range with non-negligible electron occupancy
at any reasonable temperature).  This implies that $k_{\parallel}$-dependent corrections in the DVEMA are not mandatory.

\subsection{Graded Barrier Potential}
\begin{figure}[tb]
\includegraphics*[width=0.8\columnwidth]{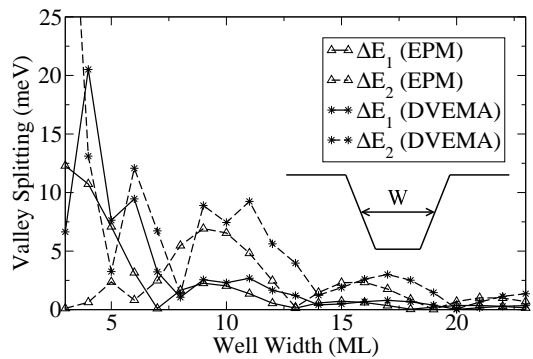}
\caption{\label{fig:SplittingGraded3Both}Valley splitting in lowest two subbands as a function of well width, W in a QW with 3-step linear graded interfaces.  The inset shows the general structure of a linear graded QW.}
\end{figure}

In real Si/SiGe quantum wells, surface segregation effects are well documented \cite{article:SurfaceScienceReportsShiraki2005}.  This refers to the ``preference'' of Ge atoms to exist on the surface of the material rather than in the bulk during molecular beam epitaxial (MBE) growth, leading to a
decrease in the magnitude of the Ge composition gradient at the nominal interfaces. Recent work \cite{unpublished:Zhang2005} has shown that atomic hydrogen etching reduces the effects of surface segregation, but this is impractical in a multilayer structure as it is extremely time consuming and surface defects caused by the etching are likely to accumulate.  It is therefore unrealistic to model a Si/SiGe quantum well as having abrupt interfaces.  The effect of graded interfaces on subband splitting is therefore considered.  The linear-graded structure shown in the inset of Fig.~\ref{fig:SplittingGraded3Both} is modelled first.  

Within the EPM calculation, the linear-graded interfaces on either side of the QW are modelled as 3-step piecewise-linear, \emph{i.e.} the interfaces spread across three\,ML, with Ge content of 17\%, 33\% and 50\% sequentially.  The results are shown in figure \ref{fig:SplittingGraded3Both}.  The well width is defined as the full-width at half-maximum (FWHM) of the envelope potential.  The results of DVEMA and EPM are in good agreement
for larger well widths --- those which allow for more than a single bound state.  The plots show that the oscillatory component of the valley splitting is unchanged, although the envelope decreases in magnitude.  This is because graded interfaces have reduced large-wave-vector Fourier components in the envelope potential, which mix the two $\Delta_{\perp}$ valleys and hence the splitting is generally smaller.

\begin{figure}[tb]
\includegraphics*[width=0.8\columnwidth]{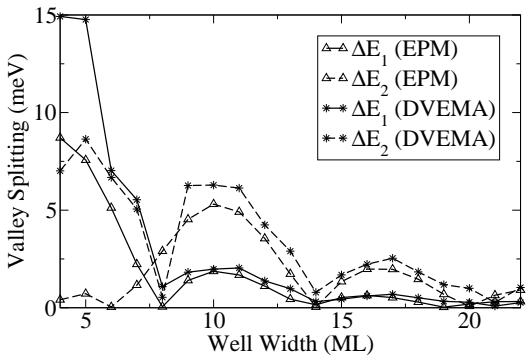}
\caption{\label{fig:SplittingGraded4Both}Valley splitting in lowest two
subbands as a function of well width in a quantum well with 4-step linear
graded interfaces.}
\end{figure}

As the width of the graded interfaces increases, the splitting is further reduced, as shown in Fig.~\ref{fig:SplittingGraded4Both} for a 4-step graded interface with Ge content of 13\%, 25\%, 38\% and 50\%, sequentially.  Again, there is a very good agreement between the two models.

\begin{figure}[tb]
\includegraphics*[width=0.8\columnwidth]{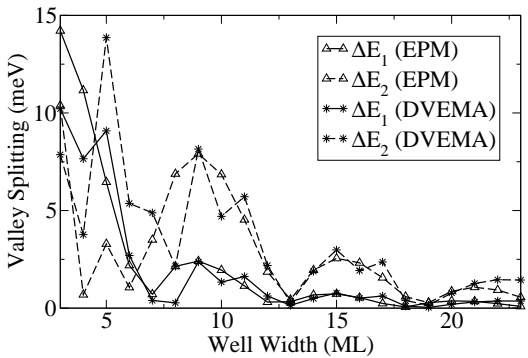}
\caption{\label{fig:SplittingGraded3NonLinBoth}Valley splitting of the lowest two subbands as a function of well width in a quantum well with 3-step non-linear graded interfaces.}
\end{figure}

A linear graded interface is a somewhat idealised model as experimental evidence shows that the interface profile is decidedly non-linear.  A 3-step grading with germanium concentrations of 13\%, 38\% and 50\% is therefore used as an approximation to a typical interface composition.  The results of EPM and DVEMA calculation are shown in Fig.~\ref{fig:SplittingGraded3NonLinBoth}.

The magnitude of the splitting is somewhat larger than for the case of linear grading, apparently because the potential gradient at the interface is now larger over a wide range of energies, thus corresponding to a
steeper linear-grading at the energies of the first and second subband minima.  The DVEMA results are again in close agreement with the EPM results.

\subsection{Double Quantum Well}
\begin{figure}[tb]
\includegraphics*[width=0.8\columnwidth]{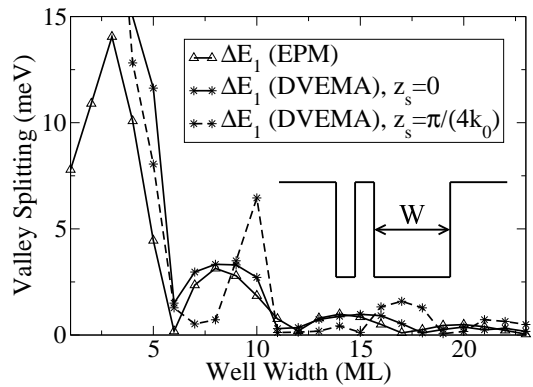}
\caption{\label{fig:splittingDoubleShift}Valley splitting in lowest subband as a function of second well width in the double quantum well structure, shown in the inset.  The results are shown for EPM and DVEMA.  In the case of the DVEMA, results are shown for two different origins for the symmetric approximation.}
\end{figure}

Next, the valley splitting in a double quantum well structure is considered.  This relatively simple structure usually provides sufficient design freedom for the required subband spacing in an optically pumped intersubband laser.  Such a structure is also a good test of the validity of the DVEMA described above, as it is asymmetric and the simple cosine modulated splitting envelope potential is no longer strictly applicable.  The simulated structure, shown in the inset of Fig.~\ref{fig:splittingDoubleShift}, has a fixed 1\,ML well separated from the second well by a 1\,ML, 50\% Ge barrier.  All other parameters are unchanged.

The results for the EPM and DVEMA calculations are shown in Fig.~\ref{fig:splittingDoubleShift}.  In this case, the structure is assumed to deviate only slightly from the square well, and therefore the periodic structure may be considered approximately symmetric about the $z=0$ position (\emph{i.e.} the left-hand side of the structure shown in the inset of Fig.~\ref{fig:splittingDoubleShift}).  The axis of symmetry, $z_s$ is therefore set at this point.  As the structure only contains a relatively small perturbation from a symmetric quantum well, the DVEMA and EPM results are still in good agreement.  The splitting energy is again lower than the simple square well case, since the left hand side (with a thin well and a thin barrier) can be viewed as a ``soft'', nonabrupt interface.  Fig.~\ref{fig:splittingDoubleShift} also shows the DVEMA results when the axis of symmetry is shifted to $z_s=\frac{\pi}{4k_0}$ such that the splitting potential becomes sine-modulated as opposed to cosine-modulated.  This represents worst-case selection of the axis of symmetry, if $z_s=0$ is assumed to be the best.  The oscillatory component of the valley splitting now appears out of phase with the EPM results, although the envelope of the oscillations is approximately correct.  The symmetric approximation is therefore dependent on the origin of the coordinate system.  However, a good estimate of the range of the valley splitting is possible, even with a poor choice of origin.

\subsection{Intersubband optical transitions}
\begin{figure}[tb]
\includegraphics*[width=0.8\columnwidth]{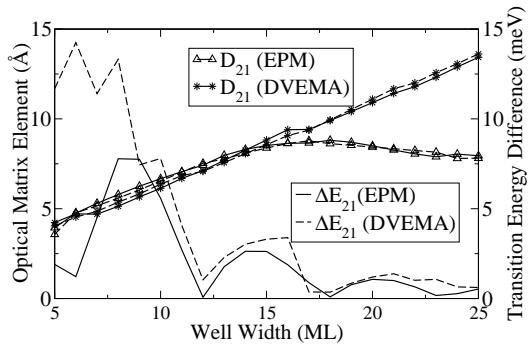}
\caption{\label{fig:MatElPlusLW}Optical matrix elements as a function of well width for the permitted transitions between the first and second subbands in the finite square quantum well considered in section \ref{scn:FiniteSquareWell}.  The difference between the two permitted transition energies is also displayed.}
\end{figure}

\begin{figure}[tb]
\includegraphics*{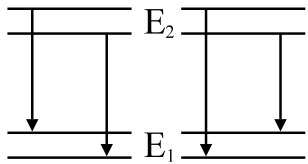}
\caption{\label{fig:PermittedStates}Permitted optical transitions in a square quantum well are usually between the two upper or lower valley-split states (left).  When close to splitting minima however, this situation is sometimes reversed in EPM simulation (right).}
\end{figure}

Optical matrix elements were calculated for intersubband transitions in the square well (section \ref{scn:FiniteSquareWell}).  The results from the EPM and DVEMA simulations are shown in figure \ref{fig:MatElPlusLW} along with the separation of the transition energies.  The difference between the optical matrix elements is small and approaches zero as the transition energies converge.  This implies a similar magnitude of spectral contribution from each pair of valley-split states.  The two methods are in close agreement for lower well widths, with the DVEMA predicting larger matrix elements at higher widths.  In most cases, when considering valley splitting of states, the permitted optical transitions are from the upper ``excited state'' to the upper ``ground state'' and between the two lower states.  However, when close to the splitting minima (at well widths of 17\,ML and 23\,ML), the converse situation sometimes applies with the EPM (fig.~\ref{fig:PermittedStates}).  The DVEMA always finds transitions to be of upper$\to$upper and lower$\to$lower character.  Transitions exhibit linewidth broadening by interface roughness and carrier scattering, typically of the order 5--10\,meV.  In the majority of cases, valley splitting is relatively small and will only cause an increase in linewidth broadening by the amount shown in figure \ref{fig:MatElPlusLW}.  However, when the valley splitting is large (for example at around 8\,ML well width), a transition line doublet may become apparent.

\section{Conclusion}
The DVEMA method presented by Ting and Chang \cite{article:PRBTing1988} has been extended to model intervalley-mixing in any symmetric structure.  DVEMA and EPM methods have been used to calculate $\Delta$-valley subband splitting in a range of symmetric and asymmetric Si/SiGe heterostructures, with both abrupt and graded interfaces.  This provides a much closer approximation to intersubband optical device structures than has been achieved previously.  The results of the two methods are in good or reasonably good agreement with each other as well as with published tight-binding results, with DVEMA demanding less than 0.5\% the computational run-time of EPM.  Subband splitting of the order of 10\,meV was predicted for abrupt-interface square quantum wells in the range of well widths of interest for silicon intersubband devices \cite{article:JCGZhang2005, article:PhysicaEPaul2003, article:PhysicaEPaul2003_2, article:PRBKelsall2005, article:ScienceDehlinger2000, article:ThinSolidFilmsZhao2006}.  It has been shown that this will typically lead to linewidth broadening, although at narrow well-widths a transition line doublet may form, with both valley-split states contributing equally to the spectrum.  The effect of surface segregation was modelled by considering both linear and non-linear composition grading at the interfaces.  This was found to reduce the valley splitting, as it is dependent upon the potential gradient at the interfaces.  Modelling of valley splitting is therefore important for the development of the as yet unrealised n-Si/SiGe quantum cascade laser.  As these structures consist of many heterolayers, the extended DVEMA method presented in this work, is a valuable tool for rapid simulation.

An important question for future designs of n-Si/SiGe intersubband lasers is whether a fast depopulation of the lower laser state may be assisted by scattering between valley-split subbands.  Alternatively, by careful selection of well widths it may possible to design structures for which the valley splitting is minimized.

\begin{acknowledgments}
This work is supported by EPSRC Doctoral Training Allowance funding.  The authors are grateful to Dr J. Zhang, Imperial College London and Dr D. Paul, University of Cambridge for useful discussions.
\end{acknowledgments}

\bibliography{InterValley}
\end{document}